\documentstyle[]{l-aa}
\input psfig.sty
\input psfonts.sty
\input times.sty
\begin{document}
\thesaurus{11                   % A&A Section 11: galaxies
            (11.01.2;           % active galactic nuclei
	     11.02.1;           % BL Lacertae objects: general
	     03.13.6)}		% methods: statistical 		
\title{Intraday variability in x-ray selected BL Lacertae objects \thanks{ Based
on observations collected at the German-Spanish Astronomical Centre,
Calar Alto, operated by the Max-Planck-Institut f\"ur Astronomie,
Heidelberg, jointly with the Spanish National Commission for Astronomy}$^,$
\thanks{Based on observations carried out at the
European Southern Observatory, La Silla, Chile during programs 
${\rm N}^{\rm o}$ 49.B$-$0036 and ${\rm N}^{\rm o}$ 53.B$-$0036}}
\author{J. Heidt\inst{1}
 \and S. J. Wagner\inst{1}}
\institute{Landessternwarte Heidelberg, K\"onigstuhl,\\
69117 Heidelberg, Germany}
\offprints{\ \protect\\
 J.~Heidt,~E-mail:~jheidt@lsw.uni-heidelberg.de}
\date{Received 15 July 1997 / Accepted 12 August 1997}
\maketitle
\begin{abstract}

We present a study of the intraday variability behaviour of two samples
of x-ray selected BL Lac objects, the EMSS and EXOSAT samples consisting of
22 and 11 sources, respectively. In both
samples we were able to detect intraday variability in less than 40\%
of the sources only. The duty cycle (the fraction of time, when a BL Lac
object is variable) in x-ray selected BL Lac objects 
is 0.4 or less.

The typical peak-to-peak amplitudes of the variability are 10\%. Typical
time-scales and an activity parameter for our variable BL Lac objects
were inferred from structure function and autocorrelation function analyses.
In only 4 BL Lac objects we were able to measure a characteristic time-scale,
which was in the range between 1.3 and 2.7 days.

Comparison with our previous study of a complete sample of 
radio-selected BL Lac objects from the 1 Jy catalogue shows that x-ray
and radio-selected BL Lac objects differ in their duty cycle by a factor of
2 and the typical peak-to-peak amplitudes by a factor of 3. The observed
time-scales are similar. We also found that the same mechanism may be
responsible for the observed variability in the x-ray selected and 
radio-selected BL Lac objects. 

The expectations of the various schemes linking x-ray selected and
radio-selected BL Lac objects have been compared to our observations.
Consistency is found for a scenario, where
x-ray selected BL Lac objects have on average stronger magnetic fields
and are seen under relatively larger viewing angles 
than the radio-selected BL Lac objects. However, the suggestion that
x-ray selected BL Lac objects have decelerating jets and radio-selected BL
Lac objects accelerating jets can also not be ruled out. In any case, any model
which explains variability on time-scales of days must be able
to reproduce the high duty cycle in radio-selected BL Lac objects
and the factor of 2 lower duty cycle in x-ray selected BL Lac objects.

\end{abstract}
\section{Introduction}

BL Lac objects are characterised by
strong and rapid variability from radio up to $\gamma$-rays, high and
variable polarization in the radio and optical regime, as well as faint or 
absent emission lines in their spectra. Additionally, in many sources
superluminal motion has been observed. Due to
these properties it is nowadays believed that their energy output is dominated
by boosted radiation from a relativistic jet pointing almost directly
towards the observer. Those objects, whose relativistic jets are pointing
more randomly onto the sky are thought to be the Faranoff-Riley I (FR I)
radio galaxies (Urry \& Padovani, 1995 and references therein).

Whereas the classical BL Lac objects have been identified from
radio surveys, most of the newly identified BL Lac objects have been
detected in x-ray surveys. These so-called x-ray selected BL Lac objects
(XBL) differ from the classical radio-selected BL Lac objects (RBL)
in many respects. They have different $\alpha_{\rm RO} - \alpha_{\rm OX}$ 
spectral indices (Stocke et al., 1985) and possibly show different cosmological
evolution (Stickel et al., 1991, Morris et al., 1991). Additionally, 
XBL are not as strongly variable and not as highly
polarized in the optical as the RBL (Jannuzi et al., 1993, 1994), they
seem to have weaker radio cores and extended emission and are less
core dominated (Perlman \& Stocke, 1993). On the other hand, no
differences have been found among the properties of their host galaxies
(Wurtz et al., 1996) and their cluster environment at redshifts below z = 0.65
(Wurtz et al., 1997).

Since the detection of ``two'' classes of BL Lac objects, several suggestions
have been made to explain their differences and to unify both classes.
Ghisellini et al. (1989, 1993) proposed that XBL and RBL are intrinsically
the same objects differing mainly by their average
viewing angle of their jets to the observer ( $< 30^{\rm o}$
for XBL and $< 15^{\rm o}$ for RBL).
Alternatively, Giommi \& Padovani (1994) and Padovani
\& Giommi (1995) suggested that the main difference between XBL and RBL 
is their spectral energy distribution (SED), 
where XBL have their cutoff of the synchrotron spectrum at
higher frequencies (UV/x-ray regime) than the RBL (near-IR/optical regime).
Since they found transition objects (RBL which had their cutoff at
higher frequencies and vice versa) they introduced a new classification scheme,
the so-called LBL (low-energy cutoff BL Lacs) and HBL (high-energy cutoff
BL Lacs). More recently, Sambruna et al. (1996) studied the multifrequency
spectral properties of BL Lac objects and explained the differences between
XBL and RBL by a systematic change of the intrinsic physical parameters
of the jet such that XBL have higher magnetic fields/electron energies
and smaller sizes than RBL. Finally, Brinkmann et al. (1996) proposed that 
XBL and RBL are intrinsically different and either originate from different
populations or have emission conditions with different physical parameters 
(e.g. accelerating jets in RBL and decelerating jets in XBL).
In addition to the differences at lower frequencies, RBL turned out to be 
$\gamma-$ray emitters at GeV energies, while all of the TeV emitting
BL Lac sources belong to the XBL subtype. If the $\gamma-$ray emission
is due to inverse Compton scattering, this result confirms that the
electron population in XBL is dominated by higher Lorentz factors.

Intraday variability (IDV ) measurements of BL Lac objects in the optical are 
one tool to probe the physics and geometry in the innermost part of
the AGN (see Wagner \& Witzel, 1995 for a comprehensive summary). 
Since optical IDV is unaffected by extrinsic effects, such as
refractive interstellar scattering, which is important in the radio domain
or gravitational microlensing, which might only be important on time-scales 
of weeks to months, such studies investigate variations intrinsic
to the source. With the current availability of medium-sized telescopes in 
combination with highly efficient CCDs, IDV studies of BL Lac objects as
faint as 20mag are possible. 
This allowed us to observe large samples of BL Lac objects within
a reasonable time, with an adequate sampling and with errors lower than in any
other frequency regime.

Despite the fact that several groups put major efforts in 
observing BL Lac objects on time-scales from hours to months (e.g.
Fiorucci \& Tosti, 1996; Miller \& Noble, 1997; Takalo, 1997) the samples
observed are mostly chosen on the basis of the brightness of the sources,
previously well documented variability, or are restricted to specific objects. 
In most cases XBL are missing, since only a few are bright enough to
be observed with smaller telescopes.

As a consequence, the optical IDV properties of XBL are poorly constrained. 
Only two XBL - the classical XBL PKS 2155-304 and Mkn 421 - have been
the subject of several detailed studies, partly during global multifrequency
campaigns (e.g. Carini \& Miller, 1992; 
Courvoisier et al., 1995; Wagner et al., 1997; Pesce et al., 1997; Heidt
et al., 1997). These observations showed that XBL have a duty cycle close 
to unity. However, in PKS 2155-304 also a quiescent period
was observed (Heidt et al., 1997). Optical variation on shorter time-scales 
(seconds to hours) has been 
investigated by Miller \& Noble (1997) and on longer time-scales by 
Jannuzi et al. (1993, 1994) and Xie et al. (1996).

In a previous study, we determined the duty cycle, typical amplitudes and
time-scales of the complete sample of RBL from the 1 Jy catalogue (Heidt \&
Wagner, 1996). The duty cycle among the RBL is very high ($\approx 0.8$), 
with amplitudes typically reaching 30\% and time-scales between 0.5 and 3 days. 

Here we present the results of our study of the optical IDV properties
of two samples of XBL, the EMSS sample extracted by Morris et al. 
(1991) and the EXOSAT sample compiled by Giommi et al. (1991). Our goal
is to determine the duty cycle, typical amplitudes and time-scales
of XBL, to compare our results with those derived from our study
of RBL and to interpret our results in view of the competing models linking 
RBL and XBL.

This paper is organized as follows: In chapter 2 the samples are
briefly described, in chapter 3 the observations and the data reduction
are summarized. Chapter 4 contains the basic results, the statistical 
analysis is presented in chapter 5. In chapter 6 we discuss the results of 
the statistical analysis and compare the results with those derived from
our study of RBL followed by the interpretation. Our conclusions are 
summarized in chapter 7. Throughout the paper ${\rm H}_{0} = 
50\ {\rm km}\ {\rm s}^{-1}\ {\rm Mpc}^{-1}$ and ${\rm q}_{0} = 0$ is assumed.

\section{The x-ray selected samples}

\subsection{The EXOSAT sample}

The EXOSAT sample of BL Lac objects was compiled by Giommi et al. (1991) 
from the EXOSAT High Galactic Latitude Survey. The 11 objects of this sample 
were selected using the following criteria:

\begin{itemize}

\item Equivalent width of emission lines $< 5$ \AA\ (rest frame)

\item 0.2 $< \alpha_{\rm ro} < 0.55$; 0.6 $< \alpha_{\rm ox} < 1.5$ 

\item x-ray flux (0.05 $-$ 2.0 keV) $> 
10^{-12}$\ ${\rm ergs\ cm}^{-2}$\ ${\rm s}^{-1}$

\item $|$ b $|$ $> 20^{o}$

\end{itemize}

\subsection{The EMSS sample}

The EMSS sample of BL Lac objects was extracted by
Morris et al. (1991) from the Einstein Medium-Sensitivity Survey.
It contains 22 objects, which were selected according to the
following criteria:

\begin{itemize}

\item Inclusion in the EMSS sample of serendipitous x-ray sources

\item Equivalent width of emission lines $< 5$ \AA 

\item Evidence for a non-thermal continuum in the spectra (Ca II break
had a contrast of less than 25\%)

\item x-ray flux (0.3 $-$ 3.5 keV) $> 5 \times 
10^{-13}$\ ${\rm ergs\ cm}^{-2}$\ ${\rm s}^{-1}$

\item $\delta > -20^{o}$

\end{itemize}

Both samples have one object in common (1207.9+3945). Therefore, this object 
was observed during two runs. Whereas for the BL Lac objects of the 
EMSS sample redshifts are available for all sources, only 4 out of 11 BL Lac 
objects from the EXOSAT sample have published redshifts. The average 
redshift of the EMSS BL Lac objects is 0.32. When relevant, we will set a 
redshift of z = 0.3 for all BL Lac objects from the EXOSAT sample, 
whose redshift is unknown.

\section{Observations and data reduction}

The observations were carried out during 9 observing runs between
September 1992 and April 1995 at the Calar Alto Observatory in Spain,
the ESO in Chile and in Cananea, Mexico.
The telescopes were equipped with a CCD and an R filter in
order to perform relative photometry. An overview about the observing
runs is given in Table 1.

\begin{table}
\caption[]{ Observing journal}
\begin{tabular}{|l|l|}
\hline
  &   \\
Observatory & Period \\
  &   \\
\hline
  &   \\
ESO, Danish 1.5m & September 1992 \\
Calar Alto, 2.2m & February 1993 \\
Calar Alto, 2.2m & May 1993 \\
Calar Alto, 2.2m & March 1994 \\
Cananea, 2.1m    & April 1994 \\
ESO, Danish 1.5m & July 1994 \\
Calar Alto, 2.2m & August 1994 \\
Calar Alto, 2.2m & December 1994 \\
Cananea, 2.1m    & April 1995    \\
  &   \\
\hline
\end{tabular}
\end{table}

In order to create a homogeneous dataset we tried to observe each BL Lac
object during seven consecutive nights with an average sampling rate
of 2 hours. Due to various circumstances (weather, technical problems etc.),
however, the sampling and the length of our data trains for
the two samples are different. 

Each BL Lac object of the EXOSAT sample was observed on 
average 2-3 times per night during 7 nights.  Apart from EXO 1146+2455, which 
could be observed during 3 nights only, each object was observed at least 
during 5 nights. About 50\% of the EXOSAT BL Lacs have been observed at 
least during 10 nights. 

Our observations of the EMSS BL Lacs were more densely sampled. Most of
the sources have been observed 4-5 times per night for a period of 6 nights,
the shortest data train was 5 nights, with a sampling of once per night 
(MS 0922.9+7459 and MS 1534.2+0148). About 25\% of the sources 
were observed during at least 9 nights (c.f. Table 2).

The integration times varied  between 5 and 30
minutes depending on the brightness of the BL Lac object, the observing
conditions and the size of the telescope. They were chosen such
that the signal to noise in the central pixel of the BL Lac object was as
high as possible but the count rate approximately 30\% below the
saturation limit or the non-linearity limit and  the count rates  of the
comparison stars varied no more than a factor of two within each campaign.

The CCD frames were reduced in a standard manner (bias subtracted, 
corrected for dark current, if necessary, 
pixel-to-pixel variations removed using twilight flat-fields).

In order to carry out relative photometry the count rates of the
BL Lac object and 5 to 10 comparison stars were measured on each frame
by simulated aperture photometry. We computed the
normalized ratio for each pair  of objects over the entire campaign
to construct lightcurves. By inspecting the lightcurve of each pair,
variable stars were found and rejected from further analysis. For the
final analysis the lightcurves were used which included the BL Lac object and
the brightest, non-saturated  comparison star. The errors were estimated from
the standard deviation (1 $\sigma$) of the lightcurve of two comparison stars
as bright as or fainter than the BL Lac object. They are typically in the
order of 0.6-2\%, in worst cases up to 5\% (corresponding to 0.06mag).

\section{Variability statistics and amplitudes}

In order to derive the number of BL Lac objects displaying variability during
the observations we applied a $\chi^2$-test following Penston \& Cannon 
(1970) to each lightcurve. As in our study of the 
radio-selected BL Lac objects from 
the 1 Jy sample (Heidt \& Wagner, 1996) we choose a 
confidence level of 99.5\% as cutoff. The $\chi^2$-test
was carried out twice. First we applied this test to all lightcurves. This
gave us the fraction of the variable and the non-variable BL Lac objects.
Since we were interested in variability time-scales from several
hours to one week, we subtracted trends on longer time-scales 
by fitting a linear slope to each lightcurve of the variable objects. 
After subtraction of these slopes, the $\chi^2$-test was applied to the 
residuals.

6 out of the 11 BL Lac objects from the EXOSAT sample displayed variability
during the observations, corresponding to a fraction of 55\%. In only 4 out
of the 11 sources (36\%) we were able to detect intraday variability.
The fraction of variable EMSS BL Lac objects was lower. Here we could
detect variability in only 7 out of 22 BL Lac objects (32\%), 6 out of this 7
showed intraday variability (corresponding to 27\% for the 
EMSS sample). In Table 2 we list the both samples along with their average R
band magnitude during the observations, their redshift, the number of
observations, the maximum time separation between two observations and in
column 6 the results of the $\chi^2$-tests for a confidence level of 99.5\%. 
A ++ sign denotes the
intraday variable, a + sign variable, but on time-scales longer as the
observing period and a $-$ sign the non-variable BL Lac objects. 
For comparison we give the confidence of the $\chi^2$-test to the residuals
after subtraction of a linear slope from the lightcurves in column 7.
The final two
columns list the variability amplitudes as defined below as well as the
measurement error and information about the observing period and 
telescope used. 

We checked, whether a lower confidence level, e.g. 90\% as used by Penston
\& Cannon (1970), would alter the distribution. We ran through the 
$\chi^2$ procedure for the lower cutoff limit as outline above.
Whereas we found no change in the classification of the BL Lac objects
from the EXOSAT sample, the number of variable EMSS BL Lac objects
increased. Now, 15 out of 22 EMSS BL Lac objects (68\%) would
be classified as variable, 8 of them (36\%) would display intraday
variability. 

Regardless of the cutoff limit used, we conclude that less than
40\% of all x-ray selected BL Lac objects from our samples displayed
intraday variability during the observations.

In order to test, if the variability amplitudes are correlated with
the average brightness of the objects, we calculated the variability 
amplitudes ${\rm Amp}~=~\sqrt{({\rm A}_{\rm max}-{\rm A}_{\rm min})^2-2\sigma^2}$ 
where ${\rm A}_{\rm max}$ and ${\rm A}_{\rm min}$ are the maximum and minimum 
values of each lightcurve and $\sigma$ the measurement errors, respectively. 
They are given for each individual object in Table 2 in column 7.

The mean amplitude for the intraday variable EXOSAT sources is 9.0$\pm$3.8\%,
for the intraday variable EMSS sources 10.6$\pm$4.5\%. Including also
the sources variable on longer time-scales, the mean amplitude for the 
variable EXOSAT sources is 8.8$\pm$3.1\% and for the variable EMSS 
sources 10.0$\pm$4.3\%.

In Figure 1  we plot the average brightness of 
the BL Lac objects during the observations against the variability amplitudes. 
For comparison we have also included the "variability amplitudes" of
the non-variable sources. Obviously,
there is no correlation between variability amplitudes and average brightness.

\begin{figure}
\psfig{figure=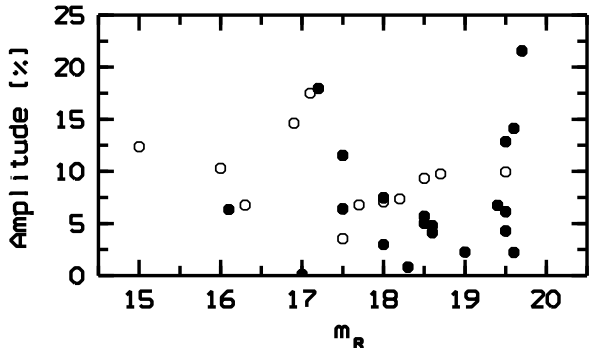,height=5cm,clip=t}
%\picplace{9.5cm}
\caption [] {Variability amplitudes versus average brightness. 
The circles correspond to the variable, the dots to the non-variable sources.}
\end{figure}

\begin{table*}
\caption[]{The x-ray selected samples}
%\begin{tabular}{|l|c|c|c|c|c|r|c|l|}
\begin{tabular}{|l|c|c|c|c|c|c|c|l|}
\hline
& & & & & & & & \\
Object & ${\rm m}_{\rm R}$ & z & n & dt [days] & $\chi^{2}$ & Confidence [\%] 
& Amplitude [\%] & Date/Telescope \\
& & & & & & & & \\
\hline
\multicolumn{9}{|c|}{}\\
\multicolumn{9}{|c|}{The EXOSAT sample}\\
\multicolumn{9}{|c|}{}\\
\hline
& & & & & & & & \\
EXO 0556.4$-$3838 & 17.5 &       & 18 & 5.20 & ++  & 99.6 & 3.5(0.6) & 09/92 (ESO,1.5)\\
EXO 0706.1+5913   & 17.0 & 0.125 & 13 & 9.87 & $-$ & 0.01 & 0.1(1.2) & 03/93 (CA,2.2) \\
EXO 0811.2+2949   & 18.5 &       & 11 & 9.80 & +   & 0.01 & 9.3(1.2) & 03/93 (CA,2.2)\\
EXO 1004.0+3509   & 19.5 &       & 13 & 9.88 & ++  & 99.9 & 9.9(1.5) & 03/93 (CA,2.2)\\
EXO 1118.0+4228   & 18.0 &       & 13 & 9.90 & $-$ & 79.0 & 7.5(2.5) & 03/93 (CA,2.2)\\
EXO 1146.9+2455   & 19.0 &       &  8 & 2.12 & $-$ & 5.6  & 2.3(1.5) & 03/93 (CA,2.2)\\
EXO 1207.9+3945   & 20.0 & 0.615 &  9 & 10.00& $-$ & 3.8  & 2.2(1.8) & 03/93 (CA,2.2)\\
EXO 1215.3+3022   & 16.0 &       & 19 & 5.12 & ++  & 100.0&10.2(1.3) & 05/93 (CA,2.2)\\
EXO 1218.8+3027   & 15.0 &       & 29 & 9.16 & ++  & 100.0&12.3(0.9) & 01/92 (CA,1.2)\\
EXO 1415.6+2557   & 17.5 & 0.237 & 23 & 6.16 & $-$ & 34.1 & 6.5(1.0) & 05/93 (CA,2.2)\\
EXO 1811.7+3143   & 18.0 & 0.117 & 22 & 5.18 & +   & 3.6  & 7.1(1.3) & 05/93 (CA,2.2)\\
& & & & & & & & \\
\hline
\multicolumn{9}{|c|}{}\\
\multicolumn{9}{|c|}{The EMSS sample}\\
\multicolumn{9}{|c|}{}\\
\hline
& & & & & & & & \\
MS 0122.1+0903 & 19.5 & 0.339 & 25 & 5.28 & $-$ & 64.9 & 6.1(1.7) & 12/94 (CA,2.2) \\
MS 0158.5+0019 & 18.5 & 0.299 & 15 & 4.20 & $-$ & 45.0 & 5.0(1.5) & 09/94 (ESO,1.5) \\
MS 0205.7+3509 & 18.6 & 0.318 & 33 & 5.33 & $-$ & 4.5  & 4.8(1.5) & 12/94 (CA,2.2) \\
MS 0257.9+3429 & 18.0 & 0.247 & 35 & 5.36 & ++  & 99.7 & 7.4(1.1) & 12/94 (CA,2.2) \\
MS 0317.0+1834 & 17.7 & 0.190 & 37 & 5.36 & +   & 63.8 & 6.8(1.1) & 12/94 (CA,2.2) \\
MS 0419.3+1943 & 19.7 & 0.512 & 36 & 5.38 & $-$ & 87.7 & 21.6(4.0) & 12/94 (CA,2.2) \\
MS 0607.9+7108 & 18.2 & 0.267 & 24 & 4.35 & $-$ & 97.0 & 7.4(2.5) & 03/94 (CA,2.2) \\
MS 0737.9+7441 & 18.0 & 0.315 & 24 & 4.33 & $-$ & 0.01 & 3.0(1.2) & 03/94 (CA,2.2) \\
MS 0922.9+7459 & 19.5 & 0.638 &  5 & 4.08 & $-$ & 93.5 & 4.3(1.0) & 03/94 (CA,2.2) \\
MS 0950.9+4929 & 18.6 & 0.207 & 22 & 4.29 & $-$ & 89.7 & 4.1(1.0) & 03/94 (CA,2.2) \\
MS 1207.9+3945 & 19.6 & 0.615 &  7 & 5.92 & $-$ & 49.4 & 14.1(5.0) & 04/94 (Can,2.1) \\
MS 1221.8+2452 & 17.1 & 0.218 & 23 & 4.28 & ++  & 100.0& 17.5(1.2) & 03/94 (CA,2.2) \\
MS 1229.2+6430 & 16.3 & 0.164 & 23 & 4.29 & ++  & 100.0& 6.8(0.6) & 03/94 (CA,2.2) \\
MS 1235.4+6315 & 18.5 & 0.297 & 17 & 6.06 & $-$ & 6.9  & 5.6(2.5) & 04/94 (Can,2.1) \\
MS 1402.3+0416 & 16.9 & 0.200 & 19 & 4.23 & ++  & 100.0& 14.6(0.8) & 03/94 (CA,2.2) \\
MS 1407.9+5954 & 19.4 & 0.495 & 24 & 9.10 & $-$ & 76.9 & 6.7(1.5) & 08/94 (CA,2.2) \\
MS 1443.5+6349 & 19.5 & 0.299 & 22 & 8.07 & $-$ & 88.8 & 12.8(2.9) & 08/94 (CA,2.2) \\
MS 1458.8+2249 & 16.1 & 0.235 & 16 & 4.19 & ++  & 99.5 & 6.3(0.7) & 08/94 (CA,2.2) \\
MS 1534.2+0148 & 18.3 & 0.312 &  5 & 3.98 & $-$ & 5.4  & 0.8(0.8) & 04/94 (Can,2.1) \\
MS 1552.1+2020 & 17.2 & 0.222 & 25 & 7.23 & $-$ & 24.4 & 17.9(4.0) & 04/94 (Can,2.1) \\
MS 1757.7+7034 & 18.7 & 0.407 & 34 & 9.16 & ++  & 99.9 & 9.7(1.5) & 08/94 (CA,2.2) \\
MS 2143.4+0704 & 17.5 & 0.237 & 33 & 9.16 & $-$ & 74.7 & 11.5(2.0) & 08/94 (CA,2.2) \\
& & & & & & & & \\
\hline	
\end{tabular}
\medskip

\noindent Column 1 gives the name of the sources, column 2 the
average brightness (${\rm m}_{\rm R}$) during the observations and column 3 
the redshift.
In column 4 the number of observations are listed, in column 5 the maximum time
separation between two observations and in column 6 the result of the 
$\chi^2$-tests. A ++ sign denotes the intraday variable, a + sign variable,
but on time-scales longer than the observing run and a - sign the 
non-variable sources
(see text for details). The confidence of the $\chi^2$-test to the residuals
after subtraction of a linear slope from the lightcurves is given in column 7.
Column 8 contains the variability amplitude as defined
in the text. Here the 1 $\sigma$ errors are included in brackets.
Column 9 finally gives the observing dates, the observatory and the telescope
used.  ESO, 1.5 = ESO Danish 1.5m; CA, 2.2,
1.2 = Calar Alto 2.2m, 1.2m, and Can, 2.1 = Cananea 2.1m, Mexico, respectively.

\end{table*}

\section{Temporal characteristics of the variability}

The temporal characteristics of the variability have been studied using
two statistical methods, structure function analysis and autocorrelation 
function analysis. Structure function analysis is a powerful tool to measure 
time-scales from the lightcurves, especially when the data sampling is rather 
inhomogeneous. Additionally, determining the slopes of the structure 
function allows to characterize the underlying physical process.
The autocorrelation analysis was used to derive typical amplitudes per
time interval. Since we described the application of both methods for
our study of the 1 Jy sample of BL Lac objects in detail
(Heidt \& Wagner, 1996), we will here only shortly summarize our 
adopted strategy.

First order structure functions (SF) were calculated from the 
lightcurves according to Simonetti et al. (1985). The first order structure
function is defined as:

\begin{equation}
{\rm SF}(\Delta {\rm t}) = \frac{1}{{\rm N}} \sum_{{\rm i}=1}^{{\rm }N} 
({\rm f}({\rm t}_{{\rm i}}+\Delta {\rm t})-{\rm f}({\rm t}_{{\rm i}}))^2
\end{equation}

The power $\alpha$ of SF $\propto (\Delta {\rm t})^{\alpha}$ characterizes
the variability and hence the underlying process.
If $\alpha$ = 1 shot noise dominates, towards flatter 
slopes ($\alpha = 0$) flicker noise becomes more important (Smith et al., 
1993). Characteristic time-scales show up as maxima in the SF 
(Wagner et al., 1996). Since we were interested mainly on
the characteristics of the variability on short time-scales and since we
wanted to compare our results to those derived for the 1 Jy BL Lac objects,
we decided to study the SF($\Delta$ t) within the time interval 
$0.5~{\rm days} < \Delta {\rm t} < 5~{\rm days}$. As in Heidt \& Wagner (1996)
we binned the SF in variable intervals such that each interval contains
the same amount of differences 
$\Delta {\rm I}({\rm t}_{1},{\rm t}_{2})$ with ${\rm t}_{2} < {\rm t}_{1}$,
where the number of bins depends on the total number of data points n.

The autocorrelation function (ACF) of each variable object was computed 
following the discrete correlation function method described by
Edelson \& Krolik (1988). As in the case of the structure function 
analysis, we had to deal with a rather inhomogeneous sampling. Therefore
our binsize $\Delta$ t was chosen to be 0.2 or 0.5 days, depending on
the evenness of the measurements between and during the nights 
(see Heidt \& Wagner, 1996, section 5.2 for details). 

\subsection{Slopes and characteristic time-scales measured from the SF}

The slopes $\alpha$ of the SF were measured by fitting a $1^{\rm st}$
order polynomial using
the least-squares method to the SF in the log(SF)$-$log($\Delta$ t) plane.
Slopes were determined only in those SF, which had at least three bins in the
range between 0.5 and 5 days. This was possible for all 13 variable sources
except MS 1757+7034, where the "best fit" was not acceptable.
The resulting slopes are given in Table 3. They range from -0.6 to 1.6, 
with a mean value of 0.8 and a dispersion of 0.7. 
The two sources with negative slopes (EXO 1215.3+3022 and MS 1229.2+6430) are 
dominated by rapid variability on time-scales around one day. 
Their structure functions are shown in Figure 2.

For two sources of each sample, we were able to measure a characteristic
time-scale by identifying pronounced maxima in the log($\Delta$ t) $-$
log(SF) plane. These are EXO 1004.0+3509, EXO 1215.3+.3022 as well as 
MS 1221.8+2452 and MS 1402.3+0416. All sources show similar intrinsic 
time-scales between 1.3 and 2.7 days.

\begin{figure}
\psfig{figure=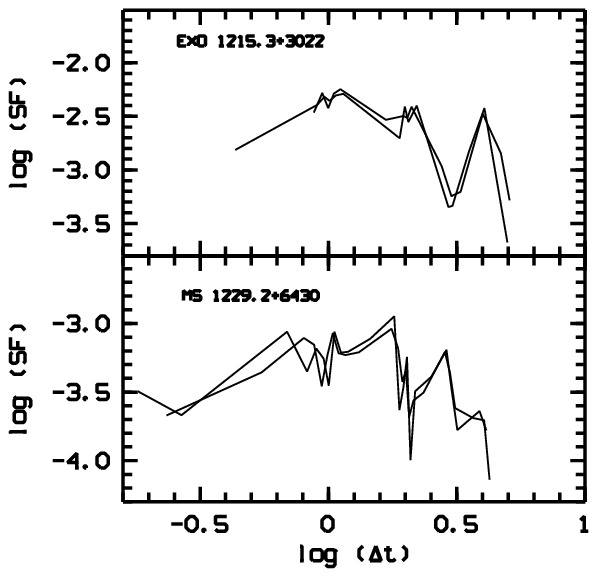,height=8cm,width=8.5cm,clip=t}
\caption[]{Structure functions of EXO 1215.3+3022 (top) and MS 1229.2+6430 
(bottom). Both sources are dominated by rapid variability on time-scales
around one day. This results in negative slopes in the range between 
0.5 and 5 days.}
\end{figure}

\begin{table}
\begin{tabular}{|l|c|c|c|}
\hline
& & &  \\
Object & Slope $\alpha$ & $\Delta {\rm t}$ [days] & $\dot{\rm I}$ [\%/day] \\
& & &  \\
\hline
& & &  \\
EXO 0556.4$-$3838 & 0.59(.12)    &          & 0.32(.49)  \\
EXO 0811.2+2949   & 1.55(.26)    &          & 0.37(.91)  \\
EXO 1004.0+3509   & 1.42(.48)    & 2.7(.27) & 0.85(.85)  \\
EXO 1215.3+3022   & $-$0.29(.13) & 1.3(.14) & 2.25(.66)  \\
EXO 1218.8+3027   & 0.59(.08)    &          & 0.86(.96)  \\
EXO 1811.7+3143   & 1.44(.22)    &          & 0.36(.57)  \\
& & &   \\
\hline
& & &   \\
MS 0257.9+3429    & 0.55(.14)    &          & 1.14(.47)  \\
MS 0317.0+1834    & 0.92(.19)    &          &            \\
MS 1221.8+2452    & 0.38(.15)    & 2.4(.25) & 4.94(1.82) \\
MS 1229.2+6430    & $-$0.63(.17) & 2.7(.25) & 2.17(.66)  \\
MS 1402.3+0416    & 1.21(.12)    &          & 2.89(2.04) \\
MS 1458.8+2249    & 1.44(.11)    &          & 1.69(.38)  \\
MS 1757.7+7034    &              &          & 1.09(.18)  \\
& & &   \\
\hline	
\end{tabular}
\caption[]{Results of the statistical analyses of the variable BL Lac
objects. Column 1 gives the name of the sources, column 2 the slopes $\alpha$
of the SF, column 3 the time-scales found  from the SF and
column 4 the activity parameter $\dot{\rm I}$ measured from the ACF. 1 $\sigma$
errors are given in brackets. The time-scales and activity parameter have been
corrected for cosmological effects.}
\end{table}

\subsubsection{The activity parameter $\dot{\rm I}$ from the autocorrelation
function analysis}

Since we want to measure a typical amplitude per time interval of
a BL Lac object, we calculated the activity parameter 
$\dot{\rm I} = \sqrt{{\rm ACF}(0)}/ \Delta {\rm t}$ 
from the ACF by fitting a $2^{\rm nd}$ order polynomial to the ACF 
of each variable object. Therefore the ACF was not normalized to unity
at $\Delta$ t = 0. ACF(0) and  $\Delta$ t were derived from the maximum 
and the decorrelation length of
the fit parabola. With the exception of MS 0317.0+1834 we could determine
$\dot{\rm I}$ for all variable BL Lac objects. The results are given
in Table 3 in column 4. Except MS 1221.8+2452, which had an $\dot{\rm I}$
of $\approx$ 5\%/day, all variable BL Lac objects
displayed a moderate variability of 3\%/day or less. 

\section{Discussion}

\subsection{IDV properties of the XBL}

Our observation show that only 6 out of 22 (27\%)
of the EMSS XBL and 4 out of 11 (36\%) EXOSAT XBL displayed IDV. Using 
a less restrict definition of IDV (cutoff limit 90\% for the $\chi^2$ test)
now 8 out of 22 (36\%) EMSS XBL would be classified as IDV, the number of
EXOSAT XBL displaying IDV would not change. Altogether, we conclude that
the detection rate is less than 40\%. Under the assumption that this
detection rate can be transformed to a duty cycle (the fraction
of time, when a BL Lac is variable), the duty cycle in XBL would be 0.4
or less.

To our knowledge no systematic study of the IDV behaviour of XBL have
been carried out so far. It is hence difficult to compare the results 
obtained with other studies. 
Xie et al. (1996) observed 6 EMSS and 1 EXOSAT XBL during
the last years with an irregular sampling, from hours to months. All sources
have been observed occasionally several times per night during two or
three subsequent nights. In 5 of the 7 sources they found 
IDV with amplitudes between 0.3 and 0.5mag on time-scales from hours to
2 days. In 3 out of these 5 sources (EXO 1218.8+3027, MS 0317+1834, MS
1402.3+0416) we found also IDV, but no XBL of our samples displayed
as large variability amplitudes as found by Xie et al. (1996).
We found mean amplitudes of $\approx 10\pm 4$\% in both samples, the most
extreme case being MS 1221.8+2452 with an amplitude of 17.5\%. 
However, the strong variability amplitudes observed by Xie et al. (1996) 
depend in all cases on one single data point in their lightcurves only, 
so their results must be regarded as uncertain.

Optical photometry and polarimetry on longer time-scales (weeks to month)
of 37 XBL have been presented by Januzzi et al. (1993, 1994). Their
sample includes 21 of the EMSS XBL (except MS 1443.5+6349) and one 
EXOSAT XBL (EXO 1415.6+2559). In spite of the faintness of some sources,
which made variability measurements impossible, they detected variability in
12 out of this 22 sources, with amplitudes of always less than 1mag.
The ``duty cycle'' on time-scales from weeks to month of this sample would 
be $\approx$ 0.55, which is slightly higher as our duty cycle of 0.4.
It is interesting to note that their duty cycle of optical polarization
(fraction of the time, where a BL Lac shows polarization $>$ 4\%)
of their sample of 37 objects is $\approx$ 0.4.

Optical Variations of 12 XBL on shorter time-scales 
(seconds to hours) have been studied by Miller \& Noble (1997). 
They found a duty cycle of $\approx$ 0.8 for the XBL they observed,
which is a factor of two higher as derived from our study on longer 
time-scales. However, they do neither quote their sources nor their
selection criteria such that a direct comparison is not possible.

From our structure function analysis we could derive characteristic time-scales
for 4 XBL only, which corresponds to a fraction of 12\%.
They were in the range between 1.3 and 2.7 days. 
The slopes of the SF range from -0.6 up to 1.6 with a 
mean of 0.8$\pm$0.7 indicating that in these sources shot noise is
the most dominating process. From our autocorrelation function analysis
we could measure a typical amplitude per time interval for all 
variable sources except MS 0317.0+1834. In all but one source (MS 1221.8+2452)
we derived an $\dot{\rm I}$ of
3\%/day which indicates that our XBL show on average modest variability only.

Two prominent XBL - the classical XBL PKS 2155-304 and Mkn 421 - were 
subject to several multifrequency campaigns during the last years
(e.g. Carini \& Miller, 1992; Courvoisier et al., 1994; Wagner et al., 1997; 
Pesce et al., 1997; Heidt et al., 1997). Whereas the observations generally 
have shown that in both XBL the duty cycle in the optical on time-scales of 
days is close to unity, in PKS 2155-304 also a quiescent period
was observed (Heidt et al., 1997). Both objects occupy the
region typical for XBL in the $\alpha_{\rm RO} - \alpha_{\rm OX}$ diagram,
however, they are unusually bright in the radio, optical and x-ray domain.
Both XBL may be an extreme member of its class.

We checked for any dependences of the time-scales,
amplitudes or the activity parameter $\dot{\rm I}$ on intrinsic properties
such as absolute magnitude or redshift and found none.

\subsection{IDV properties of XBL compared to RBL}

We compare now our results of the IDV properties of XBL 
to those derived in our study of the 1 Jy RBL sample (Heidt \& Wagner, 1996).

In Figure 3 we compare the variability statistics of the EMSS sample, the
EXOSAT sample and the 1 Jy sample. Note the clear difference on the number
of BL Lacs showing IDV, variability on time-scales longer than an observing run
(typically 7 days) and non-variable sources between XBL and RBL. 
This diagram shows that the rate of occurrence of IDV in both classes is
different. For our XBL we determined a duty cycle of $\approx$ 0.4, whereas
a duty cycle of at least 0.8 was found for the RBL 
We checked whether IDV is a coincidence
in RBL by repeated observations of various subsamples of the 1 Jy sample
and found always a duty cycle around 0.8 (Wagner et al., 1990; Bock, 1994). 
The duty cycle on time-scales of days differs between both
classes at least by a factor of 2. 

In order to quantify the difference of the rate of occurrence of IDV 
between RBL and XBL we performed a KS-test to the results of the $\chi^2$ 
distribution (see section 4). We found a probability of 
$\sim 2 \times 10^{-6}$ that the 
distributions of the $\chi^2$ have been drawn from the same parents.

Miller \& Noble (1997) determined the duty 
cycle of 12 XBL and 20 RBL. They found no difference between both classes
on shorter time-scales (seconds to hours). However, they do neither quote their
sources nor their selection criteria. This is critical for such a comparison.
Obviously, the apparent discrepancy could be understood, if the samples
of Miller \& Noble consisted mainly of XBL and RBL, which were 
already known to be very active.

\begin{figure}
\centerline{\psfig{figure=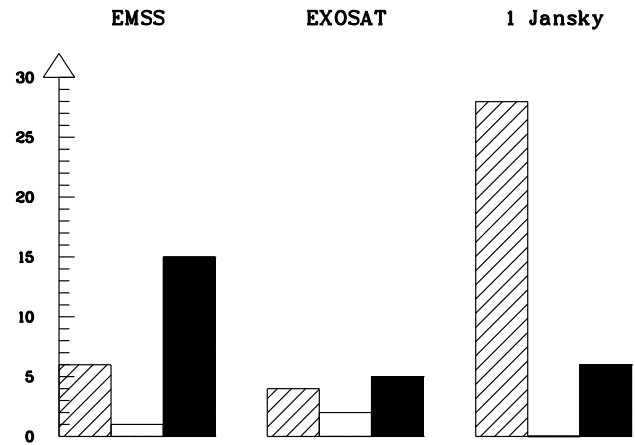,height=6cm,width=8.5cm,clip=t}}
%\picplace{4cm}
\caption [] {Histogram showing the distribution of the sources showing IDV
(shaded), the sources variable only on time-scales longer than the observing 
run (open) and the non-variable sources (black) among our three samples 
observed. Note the clear difference in the distribution between the XBL 
(EMSS, EXOSAT) and RBL (1 Jansky).}
\end{figure}

We also find significant differences in mean amplitudes between XBL and RBL.
For the intraday variable XBL we determined a mean amplitude of 10.0$\pm$4.1\% 
whereas we found mean amplitudes of 28.2$\pm$19.7\% for the variable RBL.
The amplitudes in XBL and RBL differ by a factor of three as illustrated
in Figure 4. This confirms that RBL also tend to have higher amplitudes on 
average than XBL in Miller \& Nobles microvariability study.

\begin{figure}
\psfig{figure=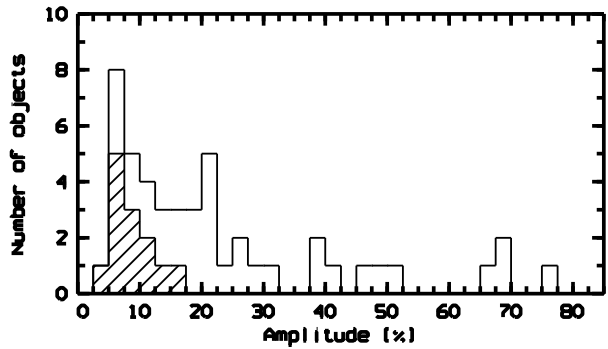,height=5cm,clip=t}
%\picplace{5cm}
\caption [] {Histogram of the amplitudes of the variable XBL (shaded)
and variable RBL (open). The difference in amplitudes can clearly be seen.}
\end{figure}

The distribution of the activity parameter $\dot{\rm I}$ (amplitude
per time interval) seems to differ among both classes. Whereas the variable 
RBL showed a wide range of $\dot{\rm I}$ up to 27\%/day,
the variable XBL had an $\dot{\rm I}$ of 3\%/day or less
(except MS 1221.8+2452 with an $\dot{\rm I}$ of $\approx$ 5\%/day).
This is shown in Figure 5, where we compare
the distribution of the $\dot{\rm I}$ found among the XBL and RBL.

\begin{figure}
\centerline{\psfig{figure=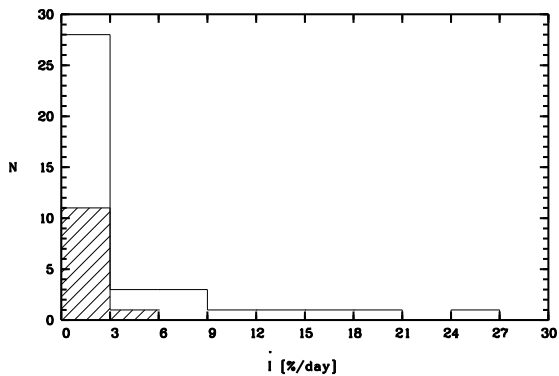,height=5cm,clip=t}}
%\picplace{5cm}
\caption [] {Histogram with the distribution of the $\dot{\rm I}$
for XBL (shaded) and RBL (open).}
\end{figure}

There also seems to be a difference between
both classes when correlating $\dot{\rm I}$ with the absolute magnitude
of the sources. Whereas we found a trend for higher $\dot{\rm I}$ 
towards higher ${\rm M}_{\rm R}$
in RBL, there is no such trend for the XBL as can be seen in Figure 6a.
However, high luminosity XBL are missing in 
our diagram. In order to test, whether
this trend is already present in low-luminosity sources,
we performed a KS-test on the distribution of the $\dot{\rm I}$ for
the XBL and RBL with ${\rm M}_{\rm R} > -27$. The cumulative distributions are
shown in Figure 6b. The KS-test gave a probability of 92.5\% that the two
distributions are drawn from the same parent population. 
As long as no information on high-luminosity XBL is available, we cannot
decide, whether only RBL show a trend of higher $\dot{\rm I}$ towards higher 
${\rm M}_{\rm R}$ or the observed
trend reflects two intrinsically different types of objects.

\begin{figure}
\centerline{\psfig{figure=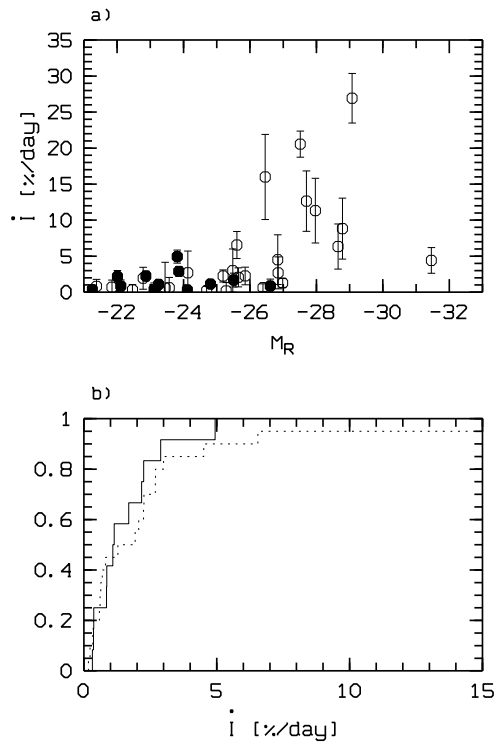,height=10cm,clip=t}}
%\picplace{5cm}
\caption [] {a) Activity parameter $\dot{\rm I}$ versus absolute magnitude
${\rm M}_{\rm R}$. The dots correspond to XBL, the open circles to RBL.
The amplitudes and time-scales involved in the derivation of $\dot{\rm I}$
have been corrected for cosmological effects. b) Cumulative distributions
of the $\dot{\rm I}$ for XBL (solid) and RBL (dashed) having 
${\rm M}_{\rm R} > -27$. The KS-test gave a probability of 92.5\% that the
two distributions have been drawn from the same parent population.}
\end{figure}

In only 4 XBL displaying IDV we could measure
a typical time-scale of the variations from our structure function analysis. 
The range found (1.3$-$2.7 days) was similar to that derived from
the 1 Jansky sample of RBL (0.5$-$3 days). 

The average slopes $\alpha$ of the SF of both classes are similar (0.8$\pm$ 0.7
for XBL and 0.8$\pm$0.6 for RBL), implying that the same mechanism might be 
responsible for the observed variability in both classes of BL Lac objects.

Finally, neither class shows a dependence of the amplitudes or time-scales
on intrinsic properties (e.g. redshift, absolute magnitude). 

\subsection{Are XBL and RBL intrinsically different?}

Our study of the IDV properties of XBL and RBL has shown that both classes
differ in many respects. The duty cycle differs by a factor of $\approx 2$,
the variability amplitudes by a factor of $\approx 3$.
Additionally, the typical amplitudes per time interval are moderate in XBL,
whereas a fraction of RBLs show also stronger amplitudes per time interval.
On the other hand, we found similar time-scales and slopes from the 
structure function analysis among both classes indicating that the 
same mechanism is responsible for the observed variability. In the following 
we will compare these results to the expectations based on the various
suggestions linking both classes of AGN.

Ghisellini \& Maraschi (1989) and Ghisellini et al. (1993) explained the
observed different overall spectral shapes of XBL and RBL by
a different viewing angle between the relativistic jet axes and the observer.
Their model is based on the assumption that the x-ray emission comes from more
closely to the base of the jet and is more isotropic than the
radio emission, which originates further away from the baseline and
is more strongly beamed. They conclude that RBL should be seen 
under an inclination angle $\theta < 15^{\rm o}$ and the XBL under an
inclination angle $\theta < 30^{\rm o}$. With this assumption and 
a Doppler factor $\delta$ scaling as $\sim 1/{\rm sin} \theta$
one would expect higher amplitudes, shorter timescales and
a higher activity parameter $\dot{\rm I}$ of the RBL as compared to
the XBL. This is confirmed by our results. 
However, based on their model, one would expect a similar duty cycle 
among both classes, which we did not found. Therefore orientation effects can 
not account for the observed IDV properties alone.

Alternatively, Giommi \& Padovani (1994) and Padovani
\& Giommi (1995) argued that a classification of BL Lac objects into
XBL and RBL is based rather on the identification in the corresponding
frequency band as a physical distinction. They proposed a new 
classification scheme, the so-called LBL (low-energy cutoff BL Lacs)
and HBL (high-energy cutoff BL Lacs), where the LBL have the high-energy cutoff
of their synchrotron spectra somewhere in the IR-optical and the HBL
somewhere in the UV/x-ray regime. Although most of the XBL are classified
as HBL and most RBL as LBL there are transition objects (XBL, which are
classified as LBL and vice versa). 
In this scenario simple predictions of IDV characteristics can be made:
since the spectral slope of the LBL is
steeper at optical frequencies as compared to the HBL one would
expect stronger amplitudes, similar time-scales and a higher 
activity parameter $\dot{\rm I}$ for the LBL as compared to the HBL
but a similar duty cycle for LBL and HBL.

In order to compare these predictions with our results, we sorted
our BL Lacs observed in LBL and HBL. All 11 EXOSAT and 22 EMSS BL Lacs
are HBL and all 34 1 Jy BL Lacs except Mkn 501 and 
PKS 2005-489, which are classified as HBL are LBL (Padovani, priv.
communication), i.e. we have now 35 HBL and 32 LBL. 
The LBL show indeed stronger amplitudes, similar time-scales and
a higher activity parameter $\dot{\rm I}$ as the HBL.
However, now 12 out of 35 HBL show IDV, corresponding to a fraction of 
$<$40\% and 26 out of 32 LBL show IDV corresponding to a fraction of 80\%. 
The duty cycle among both classes still differ by a factor of 2. Therefore
the different high-energy cutoff among both classes cannot account for
the different duty cycle. 

More recently, Sambruna et al. (1996) proposed a systematic change of the
intrinsic jet parameters as an explanation
for the observed different global spectra of XBL and RBL. 
They investigated the multifrequency 
spectral properties of the complete EMSS sample, 29 1 Jy BL Lacs and 
a complete sample of flat spectrum radio quasars (FSRQ) from the
S5 survey, which have all been observed by ROSAT in pointed mode. 
By computing composite $\alpha_{\rm ox}-\alpha_{\rm x}$ spectra
for which positive or negative values indicate concave or convex spectra,
respectively, they found that the EMSS XBL tend to have convex spectra, 
the FSRQ concave spectra and the 1 Jy RBL a mixture of both. 

Based on their results they
fitted homogeneous and inhomogeneous (accelerating) jet models
to three representative (convex, concave and intermediate
$\alpha_{\rm ox}-\alpha_{\rm x}$) multifrequency spectra and concluded that 
a different viewing angle (beaming factor) can not account for the observed 
global spectra alone, but requires instead a systematic change of intrinsic
jet parameters such as magnetic fields, maximum electron energy 
or jet size. The change should be such that objects having convex spectra
have stronger magnetic fields, higher electron energies and smaller
sizes of their jets as objects having concave spectra. 

In order to see, whether there is any trend of the shape of the composite
spectra with our IDV properties derived, we correlated their 
$\alpha_{\rm ox}-\alpha_{\rm x}$ with our amplitudes and $\dot{\rm I}$,
but found no correlation. We also calculated the mean 
$\alpha_{\rm ox}-\alpha_{\rm x}$ for our variable and non-variable BL Lacs
from the EMSS and 1 Jy sample and found an 
$\overline{\alpha_{\rm ox}-\alpha_{\rm x}} = 0.13\pm$0.61 for the
variable sources and an 
$\overline{\alpha_{\rm ox}-\alpha_{\rm x}} = -0.24\pm$0.49 for our
non-variable sources. This shows that the IDV properties are independent
of $\alpha_{\rm ox}-\alpha_{\rm x}$. 

Nevertheless, their scenario is appealing since it could reproduce all of the
IDV properties. Under the assumption that IDV is most likely originating
from perturbances within a relativistic jet
it might be possible that a varying jet configuration such as 
stronger magnetic fields in the jet of XBL
would be more resistant against such perturbances.
This could explain the different duty cycle
between XBL and RBL. Additionally, a different average viewing angle and
hence different beaming factors could also explain the different
variability amplitudes.

Different conclusions for the relationship between RBL and XBL were drawn by 
Brinkmann et al. (1996). They favor a scenario, where either RBL have 
accelerating and XBL decelerating jets or XBL are intrinsically less beamed 
BL Lac objects.  Their conclusions were drawn from a multi-frequency study of 
a large number of RBL, XBL and  highly polarized quasars (HPQ), 
which have been observed with ROSAT.
By comparing the x-ray and radio fluxes, the relations between x-ray and
radio luminosity and by correlating the flux ratios 
${\rm f}_{\rm x-ray}/{\rm f}_{\rm radio}$ and 
${\rm f}_{\rm optical}/{\rm f}_{\rm radio}$ of these three classes they
found indications that RBL and XBL are two intrinsically different types of 
BL Lac objects. 

Whereas the scenario that XBL are intrinsically less beamed objects
can not explain our results (see discussion above), 
the alternative scenario that RBL have accelerating jets and XBL have
decelerating jets does not provide unique predictions on IDV.
Additionally, the IDV properties of RBL are poorly constrained in x-rays 
(Heidt et al., in preparation) and unknown for XBL in the radio domain.\\

Obviously, the main problem for all the models discussed above
is the different duty cycle between RBL and XBL. 
One possibility to avoid this difficulty 
could be that most non-variable XBL show IDV but below our 
detection limit. However, than there must be a class of BL Lac objects
which have variability amplitudes a factor of 10-20 lower as compared 
to a typical XBL (0.5-1\% versus $\sim$ 10\%). This can hardly be tested
with the current observing techniques. 

Although several models linking RBL and XBL  
have been suggested in the last years, none of them can fully explain
the observational facts. This applies also for our observations.
Why the duty cycle among RBL and XBL differs by a factor of two
has still not been answered, especially since the physical mechanism
causing variability is most likely the same in both classes. 
Within the shock in jet model by Marscher \& Gear (1985) 
the duty cycle can be transformed in a number of shocks
travelling along a relativistic jet. Different duty cycles require 
a mechanism, which injects these shocks with a different injection rate. 

\section{Conclusions}

We have studied the variability behaviour of the 
the EMSS and EXOSAT samples of XBL on time-scales
of days. Our results can be summarized as follows:
\begin{itemize}
\item In 4 out of 11 EXOSAT sources (36\%) and in 6 out of 22 EMSS sources
(27\%) we were able to detect variability on time-scales of days or less.
This implies that the duty cycle in XBL 
is 0.4 or less.
\item In only 4 BL Lac objects we could measure a typical time-scales,
which was in the range of 1.3-2.7 days. 
\item The typical peak-to-peak amplitudes of the variability were $\sim$10\%.
\item We investigated an activity parameter $\dot{\rm I}$ from the
variable BL Lac objects. All but one BL Lac displayed an $\dot{\rm I}$
of 3\%/day or less.
\end{itemize}

Comparison with our study of the complete sample of RBL
from the 1 Jy catalogue shows that the duty cycle between XBL and RBL
differs by a factor of two, the typical peak-to-peak amplitudes
by a factor of three. Whereas the RBL showed a wide range of $\dot{\rm I}$ 
up to 27\%/day, all but one XBL had an
$\dot{\rm I}$ of 3\%/day or less. In both classes we found similar time-scales
of the variability. We also found similar slopes from our structure 
function analysis for both classes. This implies that the same process
is responsible for the observed variability. No correlation of the variability
amplitudes or time-scales with intrinsic properties have been detected
in both classes. 

We compared our results with the various suggestions linking RBL and XBL.
While we can rule out pure inclination effects or the different high-energy
cutoff of the synchrotron tail as being responsible for our observed
differences, a combination of varying jet configuration (e.g. stronger 
magnetic fields in the jets of XBL as compared to the RBL) and a different
average viewing angle might explain our observations. 
A scenario, where RBL have accelerating  and XBL decelerating jets
can not be tested with our observations. 
Nevertheless, every model explaining variability on time-scales of days
must take into account the high duty cycle in RBL and the factor of 2
lower duty cycle in XBL.

\acknowledgements{The authors would like to thank H. Bock,
A. Heines and A. Sillanp\"a\"a
for assistance during the observations and Drs. M. Camenzind and A. Witzel
for valuable and stimulating discussions. Special thanks to Dr. K. Birkle
for his permission to use a few test nights at the Calar Alto 2.2m telescope
as well as the Calar Alto OPC for the generous allocation of telescope 
time during the course of the whole project. 
This work was supported by the DFG (Sonderforschungsbereich 328).}

\end{document}